# Modeling the Dynamic Process of Inventions for Reducing Knowledge Search Costs

Haiying Ren[1,*], Yuanyuan Song [1], Rui Peng [1]

[1]*School of Economics and Management, Beijing University of Technology, Chaoyang District, Beijing, China*

[*]Corresponding author. Email: renhaiying@bjut.edu.cn

**Abstract**

A knowledge search is a key process for inventions. However, there is inadequate quantitative modeling of dynamic knowledge search processes and associated search costs. In this study, agent-based and complex network methodologies were proposed to quantitatively describe the dynamic process of knowledge search for actual inventions. Prior knowledge networks (PKNs), the search space of historical patents, were constructed, representative search rules were formulated for R&D agents, and measures for knowledge search cost were designed to serve as search objectives. Simulation results in the field of photolithographic technology show that search costs differ significantly with different search rules. Familiarity and Degree rules significantly outperform BFS, DFS and Recency rules in terms of knowledge search costs, and are less affected by the size and density of PKNs. Interestingly, there is no significant correlation between the mean and variance of search costs and patent value, indicating that high-value patents are not particularly difficult to obtain. The implications for innovation theories and R&D practices are drawn from the models and results.

## 1. Introduction

An invention is essentially a process of searching for a novel combination of concepts, methods, materials, tools, and other resources to create new knowledge (March 1991; Fleming 2001; R. Katila and Ahuja 2002; Laursen and Salter 2006). Knowledge search is considered to be a vital part of invention process, including a complex process of seeking, selection, acquisition, integration, and creation of knowledge (Huber 1991; Rosenkopf and Nerkar 2001; Riitta Katila 2002). Therefore, choosing what, where, and how to search is crucial for producing inventions with efficiency and effectiveness.

Innovation management literature has stressed that knowledge search strategies have a significant impact on innovation outcomes, for example, on whether they lead to breakthrough inventions (Fleming 2001; Kaplan and Vakili 2015; Jung and Lee 2016). Without formal definitions, search strategies are broadly understood as features of knowledge sources that can manifest in different dimensions, such as exploration vs. exploitation (March 1991), search for familiar vs. unfamiliar components (Fleming 2001; Ahuja and Lampert 2001), search for new vs. old knowledge (Nerkar 2003), local vs. boundary-spanning search (Kaplan and Vakili 2015), and original vs. ordinary search (Jung and Lee 2016). These search strategies focus on "*what or where to search*", which can provide inventors with general search targets and directions. However, in the actual R&D process, inventors often need more step-by-step direction on which knowledge elements (KEs) to search for (i.e., *how to search*), which is not available from the search strategies.

Although knowledge search can lead to innovation, it is costly. This process consumes significant resources, which can be categorized as acquisition and integration costs. Knowledge acquisition cost refers to the cost of seeking and learning new KEs; knowledge integration cost refers to the cost of recombining KEs to produce new inventions (Sidhu et al. 2007; Luo et al. 2018). As the level of knowledge search increases, the search cost may outweigh the added value of the search effort. Excessive search can reduce innovation performance, resulting in an inverted U-shaped relationship between search effort and innovation performance (R. Katila and Ahuja 2002; Phene et al. 2006; Laursen and Salter 2006; Arts and Fleming 2018). Thus, an explicit analysis of search costs can improve our understanding of knowledge searches. For practitioners, developing methods for measuring and minimizing search costs can help R&D teams achieve breakthrough innovations in less time with less cost. Despite its importance, the cost of knowledge search has not received as much research attention as the benefit of search. Although previous studies have described the cost associated with the knowledge search process, few studies have built quantitative models for measuring or reducing the cost of knowledge search in general inventive settings. This imbalance in the research must be addressed.

To measure and reduce the cost of knowledge search in inventions, both knowledge space and its dynamic search process must be modeled. Agent simulation and network science offer various dynamic models that serve similar purposes. In the literature on knowledge diffusion or transfer process, a knowledge space is modeled as a complex network composed of innovators, and innovators' deliberate search for knowledge sources is seen as a dynamic decision-making process guided by navigation rules. These rules are often referred to as "search rules" (Yu *et al.*, 2013) or "selection rules" (Ioannidis et al. 2018a, 2018b; Qiao et al. 2019), determined by evaluating other nodes based on their node attributes such as knowledge or skills (Cross et al. 2001; Borgatti and Cross 2003). Similarly, in the literature on knowledge acquisition (KA) process, the knowledge space is represented as a network consisting of KEs and their associations. One or several agents navigate this network to accumulate (i.e., search) new knowledge according to certain dynamics or search rules (da Fontoura Costa 2006; de Arruda et al. 2017; Lima et al. 2018; Zhao et al. 2018; Guerreiro et al. 2021). The efficiency of the KA process can be measured using the total number of nodes explored by the agents in a given time period (de Arruda et al. 2017; Guerreiro et al. 2021) or the total number of steps taken to explore the network (Lima et al. 2018). Different search rules lead to different learning efficiencies.

Previous research on knowledge diffusion, transfer, or acquisition has shed light on the quantitative investigation of knowledge search processes and their costs in inventions. First, complex networks enable detailed representation of KEs and dynamic simulation of knowledge search processes. Additionally, the search rules of agents govern the entire search process (how to search), enabling the tracking and measurement of knowledge search costs, while domain knowledge (Wildemuth 2004) and perception of target knowledge can influence the search rules adopted by the agents. Last, these studies show that search rules affect the amount of knowledge acquired in limited steps. Considering that any invention involves a KA process, we propose that search rules may also affect the cost of knowledge search, assuming that the cost and number of KEs are positively related.

In this study, we build an agent-based dynamic model for knowledge search processes of inventions on knowledge networks. One major goal is to investigate the effect of search rules on knowledge search costs, rather than on the total number of explored or learned knowledge in a certain time period. We would like to find good search rules that can help inventors assemble the KEs at the lowest possible cost, and are robust under search spaces of different network features such as sizes or densities. Additionally, we explore whether knowledge search cost is positively correlated with the value of inventions, which is an open question: While there is a wide speculation that high-value patents are generally more difficult to produce ("No pain, no gain"), or involving higher search risks, it has little support in the literature.

The remainder of this paper is organized as follows. In Section 2, a model for the knowledge search process of an invention is built with the knowledge network, search rules, and search cost as key components. Section 3 uses the model to simulate the search processes of 410 sample patents in photolithographic technology, and compares the search costs of the search rules. Section 4 presents the conclusions.

## 2. Model of knowledge search processes in inventions

In this research, the dynamic process of knowledge search in a specific invention setting is modeled similarly to the KA processes in knowledge networks, but with some significant modifications. First, the goal of knowledge search in KA is organizational learning, whereas the essence of invention is creative problem-solving (Altshuller 1984). Problems and solutions in patents are operationalized by problem knowledge elements (PKEs) and solution knowledge elements (SKEs), respectively. As the titles of patents generally contain the major elements of technical problems, and the abstracts contain the most important elements of the solutions, they are used to extract PKEs and SKEs, respectively. Additionally, we simplify the search process of a particular invention such that it is conducted by a single R&D agent (a team), as a small R&D team focusing on a special inventive problem often faces resource restrictions, and has to take concerted efforts. Therefore, the knowledge search process is represented by the R&D agent visiting the nodes and edges in the PKN. Finally, the knowledge search in inventions involves not only KA, but also knowledge integration (KI). In this study, a "search" action on a KE consists of its selection (by a search rule), acquisition (learning), and integration of an agent's prior knowledge. KI is conceptualized as the process of recombining newly acquired KE and prior knowledge in real-world experiments. Based on these concepts, we model the knowledge search in an invention as a process in which a R&D agent starts from its PKEs, applies a particular search rule, selects, learns, and experiments with new KEs, accumulates search costs in each step, and terminates when the "last" SKE of an invention is searched. The process resembles a jigsaw puzzle game in which the last piece of the puzzle becomes the game-winning signal.

This section describes a model for the invention knowledge search process. Sample patents ("focal patents" hereafter) in a particular technical field are used as invention cases from which the statistical properties of the cost of knowledge search are revealed. The key components of the model are the construction of prior knowledge networks, modeling of search rules, and measurement of knowledge search costs, which are detailed in the following sections.

*2.1 Constructing prior knowledge networks for focal patents*

*2.1.1 Extracting PKEs and SKEs*

This research uses patents as a data source as they contain up-to-date and reliable information concerning inventions. For a focal patent, the noun phrases from the title and abstract are proxies for its PKEs and SKEs, respectively. PKEs represent the problem to be solved by the patent, and SKEs represent the method and materials in the solution of the same patent. The PKEs and SKEs were extracted using the RegexpParser provided by NLTK. RegexpParser is a powerful parsing tool that allows users to identify specific text patterns through customized regular expressions, which is particularly adept for recognizing and extracting noun phrases. In our application, the parser

employs pattern rules defined as follows:

    <DT><VBG|VBN >*<JJ.*>*<VBG|VBN|VB>*<NN.*>*<JJ.*>*<NN.*>+;

    <VBN>*<JJ.*>*<VBN>*<NN.*>*<JJ.*>*<NN.*>+ .

*2.1.2 Collecting prior related documents*

Prior related documents (PRDs) contain technological data related to the technical problem and solution of a patent before its publication. The abstracts or titles of patents in the PRD should contain at least one PKE or SKE of the focal patent. The query (denoted by *Q*) of PRD in a patent database is: (TA: ("$PKE_1$" OR "$PKE_2$" OR …… OR "$PKE_n$") AND PBD: [* TO $d_0$]), where $d_0$ is the date before the earliest priority date of the focal patent. As the initial PRD might not cover all SKEs of the focal patent, we designed an iterative procedure for expanding the PRD as follows:

1) Let *S* be the SKEs that appear in the focal patent, but not in the initial PRD. Rank *S* in ascending order of retrievals in the patent database prior to $d_0$.
2) In each iteration, expand the initial query formula to (*Q* OR *s*), where *s* is the SKE with the smallest retrievals in *S*, update the PRD, and remove *s* from the current *S*.
3) If all SKEs in the focal patent are covered by the updated PRD, stop; otherwise, return to Step 2.

*2.1.3 Constructing prior knowledge networks*

The prior knowledge networks (PKNs) of the focal patents are constructed in three substeps:

(1) Constructing adjacency networks

In the adjacency network (AN), the nodes are KEs extracted from the abstracts of patents in the PRD. If two KEs *i* and *j* are adjacent in a sentence, they are connected to an edge. The edge weight $N_{ij}$ ($N_{ij} \in (1, +\infty]$) represents the number of times that *i* and *j* are adjacent.

(2) Constructing semantic networks

In the semantic network (SN), the nodes are the same as those in the adjacency network, whereas edge weight $S_{ij}$ is the similarity between KEs *i* and *j*. The measure of similarity takes into account factors such as the length and order of common words between the two KEs (Islam and Inkpen 2008). Analyzing the results of similarity calculation, we found that the KE pairs with similarity values above 0.7 have good semantic similarity. Thus, we selected the KE pairs with $S_{ij} \geq 0.7$ to construct the semantic network.

(3) Merging ANs and SNs into prior knowledge networks

Finally, the constructed AN and SN are merged into a PKN for the focal patent (Figure 1), which is a simple, undirected, and weighted graph. The weight value in adjacent network reflects the combination frequency of KEs in previous inventions, directly representing the correlation between KEs in R&D activities. However, the semantic relationship may be influenced by the similarity calculation method, making adjacency more significant. In the merging process, if two nodes *i* and *j* contain both $N_{ij}$ and $S_{ij}$, only $N_{ij}$ is used as the edge weight. The PKN represents the knowledge search space based on the PRDs of the focal patent. By integrating the adjacency and semantic relationship between KEs, the connectivity of the PKN and freedom of knowledge search are enhanced.

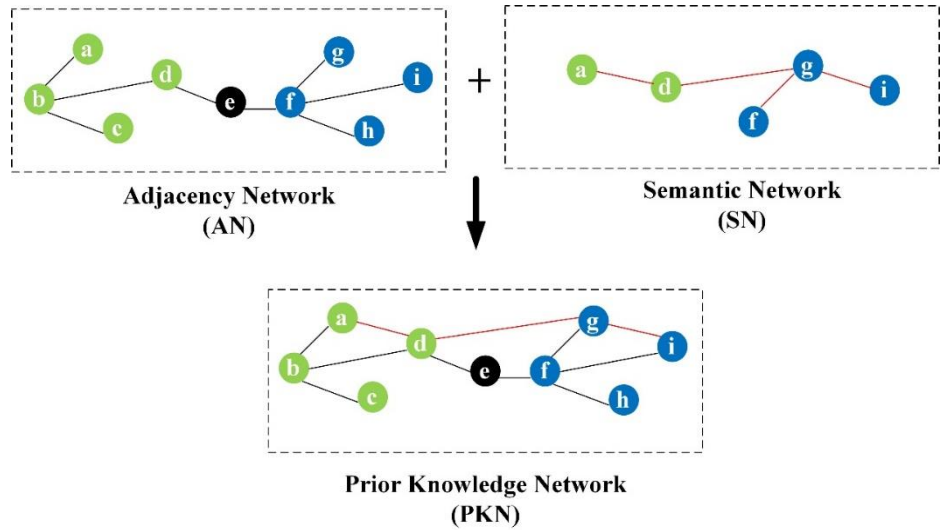

**Figure 1.** Construction process of prior knowledge network. The green nodes represent KEs of patent A, the blue nodes represent KEs of patent B, and the black nodes represent KEs shared by A and B. The black solid lines represent the adjacency relationship between KEs, and the red solid lines represent the semantic relationship ($S_{ij} \geqslant 0.7$) between KEs

Figure 2 provides a visual representation of PKN of a sample patent (US8071261B2). Within this network, the green and yellow nodes represent the PKEs and the SKEs, respectively, and the blue nodes represent the other KEs in the PKN. Red edges delineate an idealized path from PKEs to SKEs.

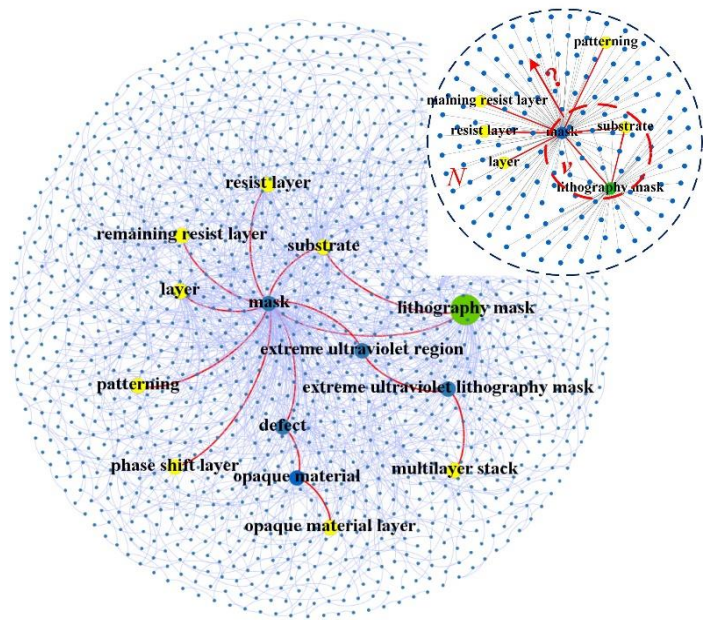

**Figure 2.** Illustration of PKEs and SKEs in the PKN and the knowledge search process of a focal patent. The red edges show the ideal (but not realistic) direct path from the PKEs to the SKEs. The enlarged picture in the upper right corner is a snippet of the knowledge search process. The three nodes in the red dotted circle are the searched nodes in a certain step of the search process. The other nodes in the black dotted circle are the neighbors of the searched nodes, serving as the candidate KEs to be selected in the next step

## 2.2 Modeling knowledge search rules

The modeling of knowledge search rules relies on the specifications of the search process. In this study, a single search rule was applied to each step of the process. The upper-right corner of Figure 2 shows the search step. Starting from the searched nodes (red dotted circle), all nodes in the black dotted circle are candidate KEs eligible to be selected and searched in the next step. Each search decision not only influences the current search cost but also has a subtle impact on further search steps, ultimately affecting the cost of knowledge search.

### 2.2.1 Uninformed search rules

An uninformed search is characterized by no prior knowledge or specification of the solution to the problem given in advance. Such search rules mimic a generic serial problem-solving approach or brute-force search (Felin et al. 2021). For example, when Edison invented the light bulb, he tested more than 6000 filament materials (Weitzman 1998). In this study, we chose breadth-first search (BFS) and depth-first search (DFS) as two common uninformed search rules.

### 2.2.2 Informed search rules

Inventors are intelligent and purposeful. Thus, they are more likely to conduct informed searches, applying some background knowledge of a PKN to select specific KEs. In this study, we assume that the R&D agent has broad knowledge of the technological field (the attributes of nodes or edges of the PKN) and uses the attributes to select and search KEs. The search process, as depicted in Figure 3, specifically as follows: First, PKEs $a, b, c$ are the searched nodes ($V$). All neighbor nodes ($N$) are evaluated by an informed search rule, and the KE with the best value ($d$) is selected and searched. Next, $d$ is moved from $N$ to $V$, and its neighbor nodes $g$ and $f$ are added to $N$. The search process continues with the updated $V$ and $N$ and terminates when the last SKE is selected and searched.

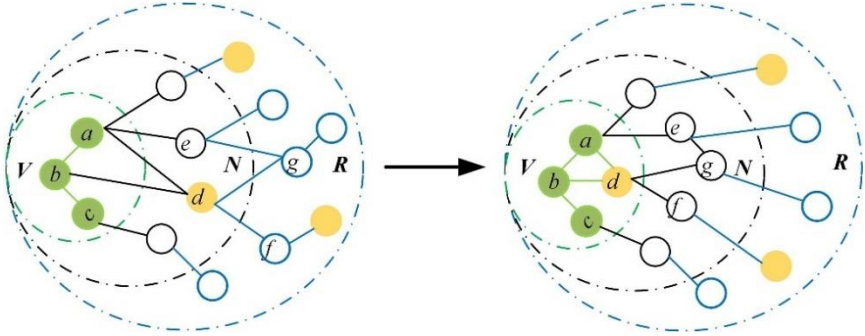

**Figure 3.** Informed search rules. $P$ represents the set of problem nodes, green nodes $a$, $b$, and $c$ represent the PKEs, and yellow nodes (such as $d$) are the SKEs

In this study, we proposed three informed search rules: Degree, Familiarity, and Recency, as shown in Table 1. The main difference between the three search rules is the attribute used to evaluate the candidate nodes.

(1) Familiarity

With the Familiarity rule, an agent gives priority to the KE in $N$ that has the largest edge weights with the searched KEs; Ties are broken by degree centrality. Its rationale comes from the fact that inventors tend to use familiar KE combinations; The more a KE combination is used in previous innovation activities, the more experience there is with the combination (Fleming 2001). Thus, an agent usually finds familiar paths (KE combinations) with high efficiency and low search costs.

(2) Degree

Using the Degree rule, an agent selects the KE in $N$ with the largest degree centrality for searching; ties are broken by the strengths of the KEs. The rationale for this rule is that the greater the degree of a KE, the greater the importance of the methods or materials it represents (Yoon et al. 2011). In addition, KEs with a large degree centrality are knowledge hubs that may facilitate finding other KEs for particular inventive problems.

(3) Recency

Unlike the Degree and Familiarity rules that select "popular" KEs or combinations, the Recency rule prefers the KE in $N$ that is in the most recent technical documents; ties are broken by the reciprocal of edge weights. Recent knowledge offers diversity and opportunities for developing innovative capacity (Cohen and Levinthal 1990). However, it may also introduce uncertainty and risk in R&D, as an agent focusing on recent knowledge may spend more time learning and experimenting (Schoenmakers and Duysters 2010).

**Table 1.** Descriptions of informed search rules. $W_{ij}$ represents the edge weight between node $i$ ($i \in N$) and neighbor node $j$ ($j \in V$); $degree_i$ denotes the degree of node $i$; $birthdate_i$ is the date when the KE first appears in a patent; $pubdate_i$ represents the publication date of the focal patent, and $strength_i$ is the strength of node $i$

| Search Rule | Evaluation Formula | Interpretation |
| --- | --- | --- |
| Familiarity | $f_i = W_{ij}$ ($i \in N, j \in V$); tiebreaker: $degree_i$ ($i \in N$) | Select the node with the maximum edge weight in the neighbor node set $N$. Ties are broken by the largest degree. |
| Degree | $f_i = degree_i$; tiebreaker: $strength_i$ ($i \in N$) | Select the node with the largest degree in the neighbor node set $N$. Ties are broken by the largest strength. |
| Recency | $f_i = pubdate_i - birthdate_i$; tiebreaker: $1/degree_i$ ($i \in N$) | Select the most recent node in the neighbor node set $N$. Ties are broken by the smallest degree. |

Degree and Familiarity rules can be viewed as models of bounded rationality (March et al. 1993) in which one must "satisfice" in the face of imperfect information and knowledge (Simon 1957). The Recency rule is often used when the conventional rules cannot yield solutions. The uninformed rules, BFS and DFS, are the most common search rules to be compared. Investigating the performance of these representative search rules with respect to their knowledge search costs is essential to understanding and improving inventive efficiency.

*2.3 Measuring the cost of knowledge search*

The proposed model concludes with the measurement, or estimation, of the cost of knowledge search, which is a major goal of the study. In this study, we defined the search cost of an invention as the time and human resources spent. Although this definition does not include the cost of the allocated financial and physical resources, it is reasonable to assume that they are positively related to the cost of knowledge search.

Any step in the knowledge search process generally consists of two actions: (1) Selecting a KE from candidate KEs in the PKN and "inserting" it into the current knowledge base, and (2) learning (acquiring) the new KE and experimenting on the ideas associated with it. In this study, we assume that the time and cost of the first action are insignificant; thus, learning and experimenting constitute most of the search cost. We further assume that the search cost is additive; that is, it can be measured by summing the learning and experimental costs of all searched KEs until all the SKEs of the invention are collected. Even so, the search cost is still difficult to measure because the time and human resources for learning and experimenting with new KEs are highly uncertain, and cost data are difficult to collect. To obtain a reasonable cost estimate, we draw on the notion of cognitive ease or cognitive strain (Kahneman 2011) which describes the mental states of less or more effort. Kahneman (2012) found that knowledge with repeated experiences can be processed with cognitive ease. It is common that when learning and undertaking tasks, the more experience one has concerning certain knowledge, the less effort is required. In the constructed PKN, we assume that the agent's experience of a newly inserted KE is mainly determined by the edge weights between the KE and current KEs based on our definitions of the edge weights. When edge weights represent the number of appearances of knowledge connections between two KEs, they are positively correlated to repeated experience and less effort for learning and experimenting. When edge weights represent the similarity between the KEs, KEs with a higher similarity normally share more common knowledge, requiring less effort. Formally, we measure the cost of searching from node $i$ ($i \in N$), a candidate KE in $N$ at step $k$ to node $j$, and a KE in the agent's current knowledge base, as the reciprocal of $W_{ij}$, i.e., the edge weight between $i$ and $j$, as in Eq. (1).

$$cost(k) = 1/W_{ij} \qquad (1)$$

As *NSN* is the total number of searched nodes for knowledge search, the total search cost (*TSC*) for invention is calculated with eq. (2):

$$TSC = \sum_{k=1}^{NSN} cost(k) \qquad (2)$$

Both *TSC* and *NSN* are used to measure the cost of a knowledge search for an invention.

**3. Simulation study in photolithographic technology**

In this section, we build simulation models of knowledge search rules and costs for a particular field, photolithographic technology, compare the search costs (*TSC* and *NSN*) under various search rules, study the impact of network size on search costs, and investigate the relationship between patent value and search costs. PatSnap (PatSnap, 2023), a comprehensive worldwide patent database, was used. The search formula is: "TAC: (lithography) AND IPC: (H01L OR G02B OR G03B OR G03F) AND PRIORITY _ DATE: [20060101 TO 20151231] AND PRIORITY _ COUNTRY: (US)"; 2447 patents were retrieved. "TAC" stands for titles, abstracts, and claims. The end year of the priority dates is 2015, as we are investigating the correlation between search cost and patent value, which is only available years after patent application. Focal patents were randomly sampled from the retrieved patents. There were a few focal patents with partial SKEs that were not in the largest connected components (LCC) of their PKNs, which means that their solutions were unsearchable. These patents were excluded, resulting in 410 focal patents. One reason for sampling was to investigate the cost of knowledge search rather than surveying the entire technological field; another reason was to reduce the computational burden of investigating all photolithographic technology patents. Based on the network construction method proposed in Section 2.1, we constructed a PKN for each focal patent.

*3.1 Cost comparison of search rules*

A major goal of the simulation study is to compare the search costs (*TSC* and *NSN*) of inventions under the representative search rules described. In Figure 4, the search costs of 410 sample patents in photolithographic technology are illustrated with violin plots, which depict the distributions of the search costs under the five rules using density curves accompanied by overlaid box plots with the means and quantiles of the distributions. In Figure 4 (a) and Figure 4 (b), the distributions are shown in ascending order of the mean *TSC* or *NSN*. It is observed that most *TSC* and *NSN* are below 10000 for Familiarity, Degree, and BFS, whereas most of the cost measures are above 10000 for DFS and Recency. In addition, it is observed that the search costs with Familiarity, Degree, and BFS rules are similar; the same is approximately true for Recency and DFS rules.

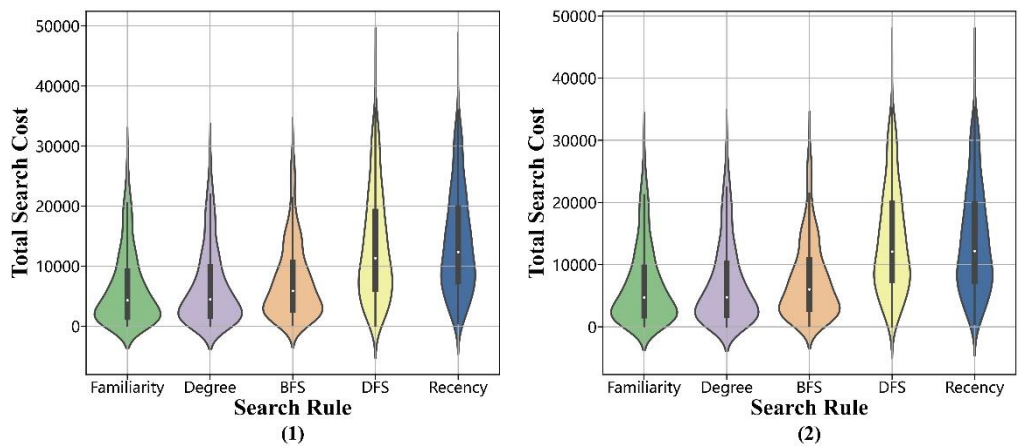

**Figure 4.** Violin plots of search costs under the five search rules

Considering that *TSC* and *NSN* under different search rules are correlated and do not conform to normal distribution, we further use the Friedman's test to compare the search costs of different search rules. The results of Friedman's test (Table 2) show that the *TSC* and *NSN* of the five search rules are significantly different ($p < 0.05$). The Post-Hoc analysis (Table 3) conducted on the adjacency search rules illustrated in Figure 4 examined the differences between them. The results show that significant differences in *TSC* between BFS and Degree, as well as between DFS and BFS. On the other hand, the differences in *TSC* between Degree and Familiarity, and between Recency and DFS, are not significant. Furthermore, the *NSN* results (not shown due to space constraints) are similar to Table 3. In summary, in terms of search cost (*TSC* and *NSN*), the five search rules can be ordered as Familiarity / Degree < BFS < DFS / Recency.

**Table 2.** *TSC*, *NSN*, and Friedman's test for the five search rules

|  | Search Rule | Median | Std. deviation | Statistical Quantities | *p*-value | Cohen's f |
|---|---|---|---|---|---|---|
| Total search cost (*TSC*) | Familiarity | 4285.573 | 6103.954 | 1030.578 | .000 | 0.46 |
|  | Degree | 4465.45 | 6335.278 |  |  |  |
|  | BFS | 5851.515 | 6225.1 |  |  |  |
|  | DFS | 11225.997 | 8927.101 |  |  |  |
|  | Recency | 12293.714 | 8509.667 |  |  |  |
| Number of searched nodes (*NSN*) | Familiarity | 4739.5 | 6294.858 | 1084.213 | .000 | 0.459 |
|  | Degree | 4775 | 6505.091 |  |  |  |
|  | BFS | 5992 | 6223.223 |  |  |  |
|  | DFS | 11981 | 8443.516 |  |  |  |
|  | Recency | 12108.5 | 8462.055 |  |  |  |

**Table 3.** Nemenyi post-hoc test following Friedman's test on *TSC*

| Pairing variable | Median ± Standard Deviation | | | Statistical Quantities | p | Cohen's d |
| --- | --- | --- | --- | --- | --- | --- |
| | Pairing 1 | Pairing 2 | Pairing Difference (Pairing 1-Pairing 2) | | | |
| Degree - Familiarity | 4465.45±6335.278 | 4285.57±6103.95 | 179.88±231.32 | 1.124 | 0.900 | 0.05 |
| BFS - Degree | 5851.52±6225.10 | 4465.45±6335.28 | 1386.07±110.18 | 7.028 | 0.001 | 0.137 |
| DFS - BFS | 11225.99±8927.10 | 5851.52±6225.10 | 5374.48±2702.00 | 25.331 | 0.001 | 0.748 |
| Recency - DFS | 12293.71±8509.67 | 11225.99±8927.10 | 1067.72±417.43 | 3.467 | 0.102 | 0.085 |

Then, we compared the dynamic performance of the five search rules. We randomly selected six patents from the sample and plotted the number of SKEs found in each step for each patent, as shown in Figure 5 where L represents the node number of the largest connected component (*LCC*). Notably, Familiarity and Degree have the best dynamic search performance; Recency and DFS have the worst. DFS and Recency tended to be much slower than the other rules in finding the first few SKEs, rushing toward the end to pick up the remaining SKEs, largely because most nodes had already been searched at this point. In contrast, the dynamic search plots for the BFS, Degree, and Familiarity search rules resemble convex curves, meaning that most SKEs are found at much lower search costs; a few difficult SKEs account for the remaining search costs.

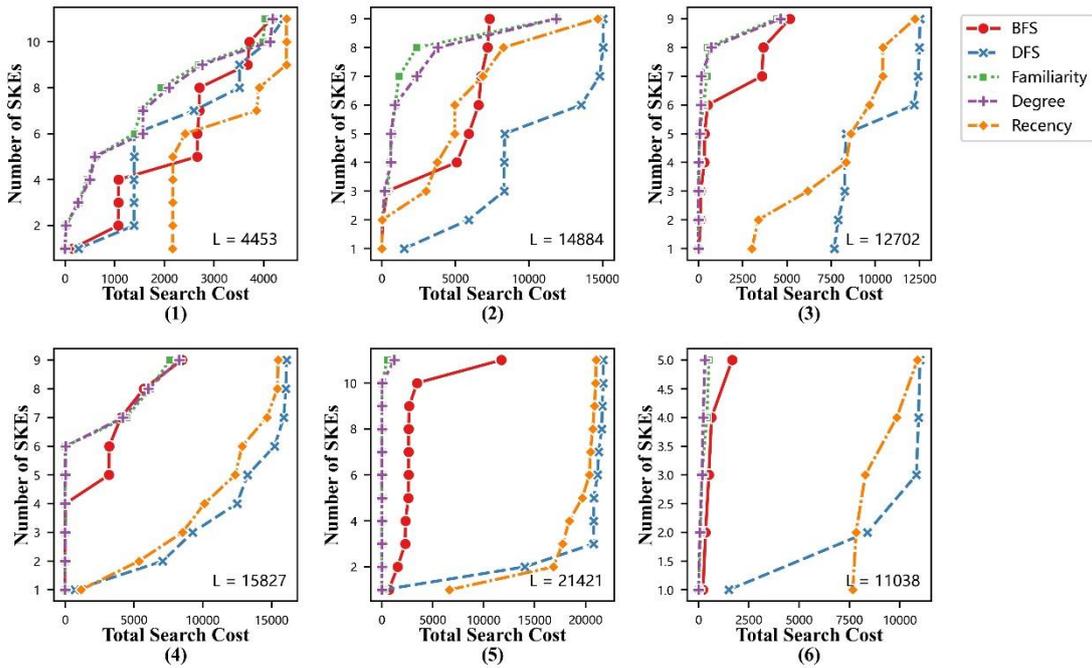

**Figure 5.** Search progress chart for different rules for six sample patents

*3.2 Effect of network size and density on search cost*

From the analysis in Section 3.1, it is found that Familiarity and Degree all demonstrate good search performance in terms of *TSC*, *NSN* and search progress. A natural question follows: Do the rules with lower search cost also have robust performance under different network sizes and densities? In this section, the impact of network size and density on *TSC* under different search rules is analyzed, where network size is measured as the number of nodes in the *LCC*. Figure 6 and Figure 7 show the relationship of *TSC* to network size and network density, respectively.

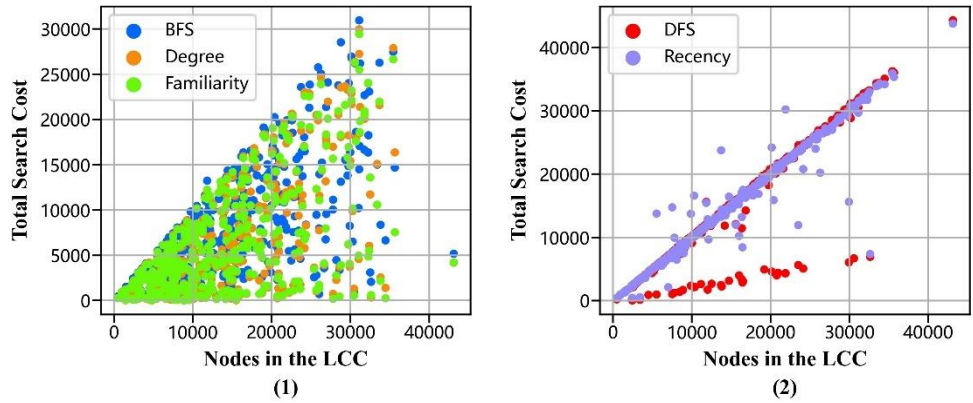

**Figure 6.** Scatter diagram of node number of LCC and *TSC*

Figure 6 shows that the network size is positively correlated with *TSC* under all five search rules. Under the Recency and DFS search rules, the relationship has a linear trend ($R^2 > 0.8$), whereas the linear relationship is much weaker for the Familiarity, Degree, and BFS search rules ($R^2 < 0.5$). This means that these rules are less affected by network size. From the results shown in Figure 7, it is apparent that the network density is negatively correlated with *TSC* under all five search rules, implying that sparce networks with less KE connections are more difficult to search. However, Familiarity, Degree, and BFS are more robust in sparce networks because they have a higher chance of obtaining inventions at low cost, while the chance is slim under the Recency and DFS rules.

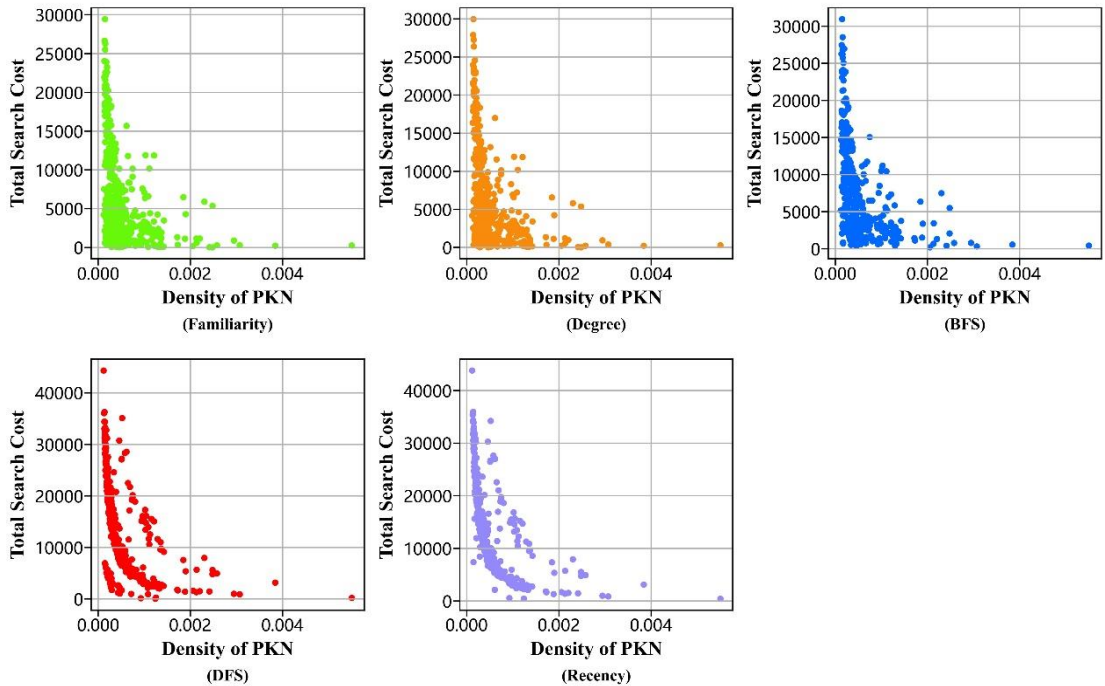

**Figure 7.** Scatter diagram of density of LCC and *TSC*

*3.3 Relationship between patent value and TSC*

An interesting question concerning the cost of knowledge search is whether it correlates with the invention value. It is commonly believed that the higher the value a patent has, the more difficult it is to find its KEs. However, our simulation results do not support this belief. In our simulations, we used the 5-year forward citation count to measure the value of a patent; previous research has shown that forward citation is a good proxy for technological and economic value, including the generated consumer surplus (Trajtenberg 1990), expert evaluation of patent value (Albert et al. 1991), and patent renewal rate (Harhoff et al. 1999). Table 4 shows the Kruskal-Wallis's test of *TSC* for patents of different value types under different search rules; patents are categorized by *citations* (5-year citation counts) as zero-cited, medium-cited (1 ≤ *citations* ≤ 19), and highly cited (*citations* ⩾ 20). The p-values of Kruskal-Wallis's test are all greater than 0.05. Thus, there is no evidence that the mean *TSCs* of patents of different value types are significantly different, as a group or pairwise. This implies that more valuable inventions are not more difficult to obtain than ordinary ones.

Table 4. *TSC* statistics and Kruskal-Wallis's test for patents of different value types

| Search Rule | Patent value type | N | Mean | Std. Deviation | $\chi^2$ | *p*-value |
|---|---|---|---|---|---|---|
| BFS | Zero-cited | 119 | 7072.690 | 6217.910 | 2.133 | 0.344 |
| | Medium-cited | 254 | 7562.951 | 6119.175 | | |
| | Highly cited | 37 | 8369.789 | 6957.502 | | |
| DFS | Zero-cited | 119 | 13650.428 | 9630.905 | 1.847 | 0.267 |
| | Medium-cited | 254 | 12833.849 | 8590.846 | | |
| | Highly cited | 37 | 14782.651 | 8723.833 | | |
| Familiarity | Zero-cited | 119 | 5587.129 | 5750.656 | 3.470 | 0.176 |
| | Medium-cited | 254 | 6694.477 | 6326.615 | | |
| | Highly cited | 37 | 5874.558 | 5551.369 | | |
| Degree | Zero-cited | 119 | 5886.702 | 5872.834 | 2.639 | 0.397 |
| | Medium-cited | 254 | 6963.066 | 6551.521 | | |
| | Highly cited | 37 | 6532.955 | 6221.612 | | |
| Recency | Zero-cited | 119 | 14270.646 | 9347.325 | 1.025 | 0.599 |
| | Medium-cited | 254 | 13700.093 | 8135.854 | | |
| | Highly cited | 37 | 15020.932 | 8130.788 | | |

This result is surprising and counterintuitive. However, many cases in the history of technology show that successful inventions need not be difficult to search. For example, we consider the invention of Post-it notes. In 1968, Spencer Silver at 3M attempted to develop a super-strong adhesive. He accidentally invented a non-sticky adhesive that 3M did not find useful. Arthur Fry inadvertently used this "failure" to make a peelable marker for his hymnal book, and 3M Post-it notes were conceived (Davis et al. 2013). Post-it notes are not any *easier* than other adhesives, with "repeated uses" (an uncommon element) in its SKE. Even so, the knowledge search path for Post-it notes is almost the same as that for other 3M adhesives; thus, the search cost is similar. This case is instructive for R&D activities. Essentially, inventors need not excessively seek difficult technology. Finding wider and more meaningful applications for inventive problems and using better search rules are often more productive and cost-efficient approaches than using extremely challenging techniques.

There is a common belief that patents with higher values generally lead to higher risk that can manifest as higher variations in search costs. However, Figure 8 does not support this belief. In Figure 8 the variances of *TSC* for the three patent categories are similar for all search rules, demonstrated by F-test results (BFS: $F = 0.561$, $p = 0.571$; DFS: $F = 1.555$, $p = 0.212$, Familiarity: $F = 1.986$, $p = 0.131$; Degree: $F = 1.834$, $p = 0.161$, Recency: $F = 2.157$, $p = 0.117$). Under some rules, highly cited patents had lower standard deviations than the other categories. The Familiarity rule had the lowest mean standard deviation of all search rules for all three patent categories. Thus, it is suggested that high-value patents do not necessarily have higher risks in terms of search costs, and smart search rules can effectively lower the risks.

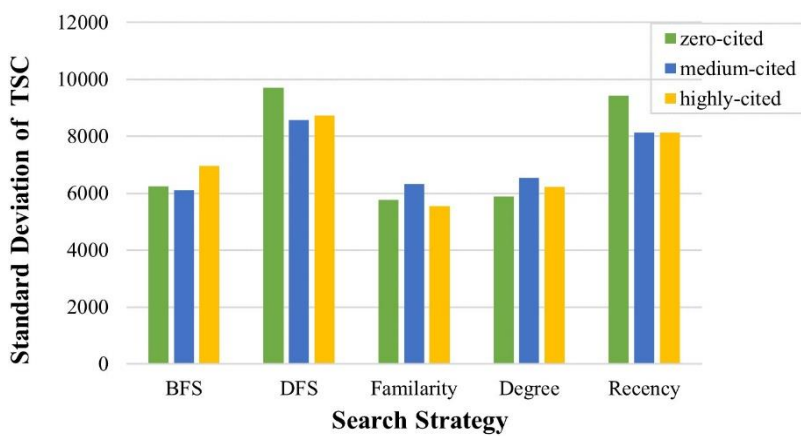

**Figure 8.** Standard deviations of *TSC* for different patent value types

**4. Conclusion**

A knowledge search is a key process for inventions. Existing literature delves deeply into knowledge search strategies and shows their impact on innovation performance. However, there is a dearth of research on both the dynamic process of invention and the measurements and determinants of the cost of a knowledge search. In this study, we modeled the dynamic process of knowledge search for inventions based on agent simulation and complex networks, built prior knowledge networks (PKNs) for focal inventions, formulated basic knowledge search rules, and measured search costs for inventive problem-solving. Simulation experiments with historical patents in photolithographic technology showed that Familiarity and Degree search rules outperform BFS, DFS, and Recency in both search cost and progress. These rules also exhibited greater robustness, being less sensitive to changes in the network environment. Additionally, we found that high-value patents do not necessarily have higher mean search costs or search risks (variations in costs) than other patents under any search rule.

This study has theoretical and practical implications. In addition to the benefits it brings, knowledge search also incurs costs. To measure knowledge search costs, it is essential to model the dynamic process of knowledge search. However, existing literature focuses on *what* or *where to search* (search strategies), but few model *how to search* through dynamic inventive processes (search rules). This is an important reason for the lack of quantitative investigation of knowledge search costs. This study is among the first to model the dynamic processes of knowledge search for inventions. It quantitatively constructs the search space, formulates the search rules of an R&D agent, and estimates the costs in the context of inventions. Therefore, it makes a theoretical contribution to the study of inventions.

In R&D practices, our simulation results show that search rules, which determine the selection of KEs to search for next, greatly impact the cost of invention. Among these rules, searching for familiar or popular KEs is more cost-effective, as agreed upon by most inventors. Such knowledge requires less effort to acquire and experiment with while offering higher reliability in associated technical methods (Ahuja and Lampert 2001). Another instructive conclusion is that more valuable inventions are not necessarily more "difficult" to obtain, nor do they have significantly higher risks in knowledge search. Thus, our advice to inventors is: "Think hard about the usefulness of your invention, rather than the complexity of your solution." Methodologically, inventors can model the dynamic search process for inventive problems with knowledge networks and use the suggested search rules (or develop more suitable rules) to guide knowledge search in their technological field, which should reduce search costs and effort.

Our study has several limitations. First, the PKNs, which represent the search space of R&D agents, were constructed in a simplistic manner by iteratively checking and inserting SKEs as if they were known in advance. Although these PKNs still enable fair comparison of search costs under different search rules, they may not fully describe the actual complex construction processes of PKNs. In the future, more realistic PKNs could be constructed. Moreover, a real-world inventor rarely sticks to any single uninformed or informed rule throughout the knowledge search process; instead, he or she tends to switch rules and adapt to different conditions. Such "adaptive rules" should be formulated and compared in the future. Additionally, this study can be extended to more general cases in which multiple R&D agents conduct complex cooperative knowledge search. Finally, measurement of knowledge search cost is by no means perfect. In the future, the additivity assumption of search cost may be relaxed to better approximate the true knowledge search cost. Digitalization of enterprise R&D processes may provide the necessary fine-grained research data for modeling PKNs, search rules, and search costs in a more realistic setting.